# Decoding Dopant-Induced Electronic Modulation in Graphene via Region-Resolved Machine Learning of XANES

Yinan Wang[1,2,3], Arpita Varadwaj[1*], Teruyasu Mizoguchi[2], Masato Kotsugi[1]

[1] Department of Material Science and Technology, Tokyo University of Science, 6-3-1, Niijuku, Katsushika, Tokyo, 125-8585, Japan

[2] Institute of Industrial Science, The University of Tokyo, 4-6-1, Komaba, Meguro, Tokyo 153-8505, Japan

[3] Department of Materials Engineering, the University of Tokyo, 7-3-1, Hongo, Bunkyo, Tokyo, 113-8656, Japan

*Corresponding author's e-mail: varadwaj.arpita@gmail.com

**Abstract:**

Revealing how heteroatom doping alters the local electronic structure of graphene is crucial for understanding and controlling its functional properties. In this study, we combine density functional theory (DFT) and machine learning (ML) to interpret how boron (B) and nitrogen (N) dopants influence the local electronic environments of graphene. A dataset of 415 DFT-simulated XANES spectra from 91 distinct configurations was analyzed using a region-specific approach by decomposing each spectrum into π*, σ*, and post-edge regions. Random forest models trained on these spectral segments identified the π* region as the most informative for predicting key local electronic descriptors, particularly the Bader charge and mean dopant-carbon bond length. The Bader charge quantifies dopant-induced charge redistribution and local bonding polarity, directly reflecting the degree of electronic perturbation introduced by heteroatom substitution. The enhanced predictive power of the π* region arises from its strong coupling to the perturbed π-electron network, which captures these charge-transfer and hybridization effects more effectively than σ* or post-edge regions. These findings establish Bader charge as a robust and physically meaningful descriptor for quantifying dopant-induced electronic modulation and demonstrate that region-resolved ML analysis of XANES spectra provides a powerful pathway to uncover structure–property relationships in doped graphene and related materials.

## 1. Introduction

Graphene, a two-dimensional monolayer of carbon atoms arranged in a honeycomb lattice, has attracted tremendous attention due to its exceptional carrier mobility, mechanical strength, and thermal conductivity.[1,2] Its linear Dirac-cone dispersion enables massless charge carriers, making it an appealing platform for next-generation nanoelectronic and energy-storage devices.[3] However, pristine graphene lacks an intrinsic bandgap, limiting its applicability in semiconductor technologies.[4] Chemical doping with heteroatoms such as boron (B) and nitrogen (N) has therefore emerged as an effective strategy to tune its electronic structure by introducing charge carriers, defect states, and modified bonding environments. Substitutional doping perturbs the local π-conjugated network, altering charge distribution and hybridization in a manner that depends strongly on dopant identity, concentration, and lattice position.[5,6] Boron, being electron-deficient relative to carbon, typically induces p-type behavior and generates electron-accepting sites that enhance adsorption and catalytic activity in electrochemical reactions. In contrast, nitrogen acts as an electron donor, leading to n-type modulation and localized charge accumulation that can significantly influence reaction pathways and transport properties. These dopant-induced perturbations of the π-electron network directly impact the density of states near the Fermi level and the chemical reactivity of graphene, making precise characterization of local charge redistribution essential for rational material design. The resulting functionality whether enhanced conductivity, catalytic activity, or chemical reactivity depends sensitively on dopant type, local coordination, and charge-transfer dynamics, underscoring the need for quantitative descriptors that directly connect atomic structure to local electronic behavior. [7,8]

X-ray Absorption Near-Edge Structure (XANES) spectroscopy provides an element-specific probe of local electronic environments and unoccupied electronic states. In graphene and related carbon materials, the carbon K-edge directly probes transitions from the C 1s core level to unoccupied π* and σ* states, offering direct insight into hybridization, symmetry breaking, and perturbations of the delocalized π-electron network. Experimental studies have shown that heteroatom doping, including BN incorporation, can induce measurable bandgap opening and π–π* band modification in graphene, with corresponding changes observed in XAS and XES spectra.[9,10] These findings highlight that dopant-induced charge redistribution and local bonding distortions are directly encoded in the carbon K-edge spectral features.

Because the electronic properties of graphene, such as carrier transport and bandgap modulation are governed by subtle variations in the π-system, carbon K-edge XANES serves as a powerful diagnostic tool for monitoring electronic restructuring.[11–13] However, translating spectral variations into quantitative structural or electronic descriptors remains challenging. Conventional analysis often relies on peak assignments or qualitative comparison with reference spectra, which may overlook complex and nonlinear correlations between spectral regions and local electronic properties.

In recent years, ML has emerged as a powerful approach for data-driven interpretation of spectroscopic data. [14–16] XANES spectra consist of high-dimensional, highly correlated features whose variations may arise from subtle and simultaneous changes in geometry, coordination, and charge redistribution. Disentangling these intertwined effects using conventional peak-based analysis is often non-trivial, particularly when multiple dopant configurations or concentrations coexist, as is common in experimentally synthesized graphene materials. By learning complex, nonlinear relationships between spectral features and atomic-scale properties, ML models enable systematic extraction of structural and electronic descriptors directly from spectral data.

For instance, Timoshenko *et al.* demonstrated neural-network-based prediction of coordination environments from XANES, while Torrisi *et al.* employed random-forest models to predict Bader charge and mean nearest-neighbor bond length in transition-metal oxides using multiscale polynomial features.[17,18] Chen *et al.* trained a feedforward neural network on C K-edge XANES spectra of organic molecules to predict ground-state C s- and p-orbital PDOS in both occupied and unoccupied states. [19] While such studies highlight ML's potential to accelerate XAS interpretation, most treat the spectrum as a single, undifferentiated input. A few studies have begun to probe the spectral origin of predictive power: Fujikata *et al.* showed through systematic sub-window analysis of Si K-edge XANES that edge-region features encode valence information while structural recovery requires the full spectral range.[20] However, a physically motivated, region-resolved framework, one in which spectral decomposition is guided by the distinct electronic transitions underlying each region, remains largely unexplored, particularly for doped carbon systems where charge transfer is subtle yet critically influences electronic structure.

Moreover, the role of Bader charge a physically grounded measure of charge redistribution derived from the topology of the electron density as a quantitative descriptor linking local bonding perturbations to XANES response remains largely unexplored for doped carbon systems, where charge transfer is subtle yet critically influences electronic structure.[21]

Towards this end, in this work, we develop a region-resolved machine-learning framework for analyzing carbon K-edge XANES spectra of B- and N-doped graphene. Using a systematically constructed set of DFT-generated spectra spanning diverse B and N dopant concentrations and configurations, we establish a comprehensive spectral dataset. Each spectrum is subsequently decomposed into physically meaningful regions (such as, $\pi^*$, $\sigma^*$, post-edge, and full spectrum) and evaluated using a random forest regression model. We demonstrate that the spectral features encode sufficient information to quantitatively predict both a structural descriptor (dopant–carbon bond distance) and an electronic descriptor (dopant charge). To identify an electronic descriptor capable of capturing dopant-induced charge redistribution, we evaluated commonly used population analysis schemes. Given the predominantly covalent nature of the graphene network and the localized perturbations introduced by substitutional dopants, Bader charge analysis was selected due to its zero-flux partitioning of the total electron density, which enables spatially resolved characterization of bonding-induced charge modulation. This charge redistribution directly influences the local electronic structure and associated spectral signatures, establishing a transparent link between XANES features and bonding perturbations. The resulting framework enables XANES to serve not only as a fingerprinting tool but as a quantitative probe of local electronic and geometric variation, offering a pathway for data-driven screening and electronic structure engineering in functional carbon materials.

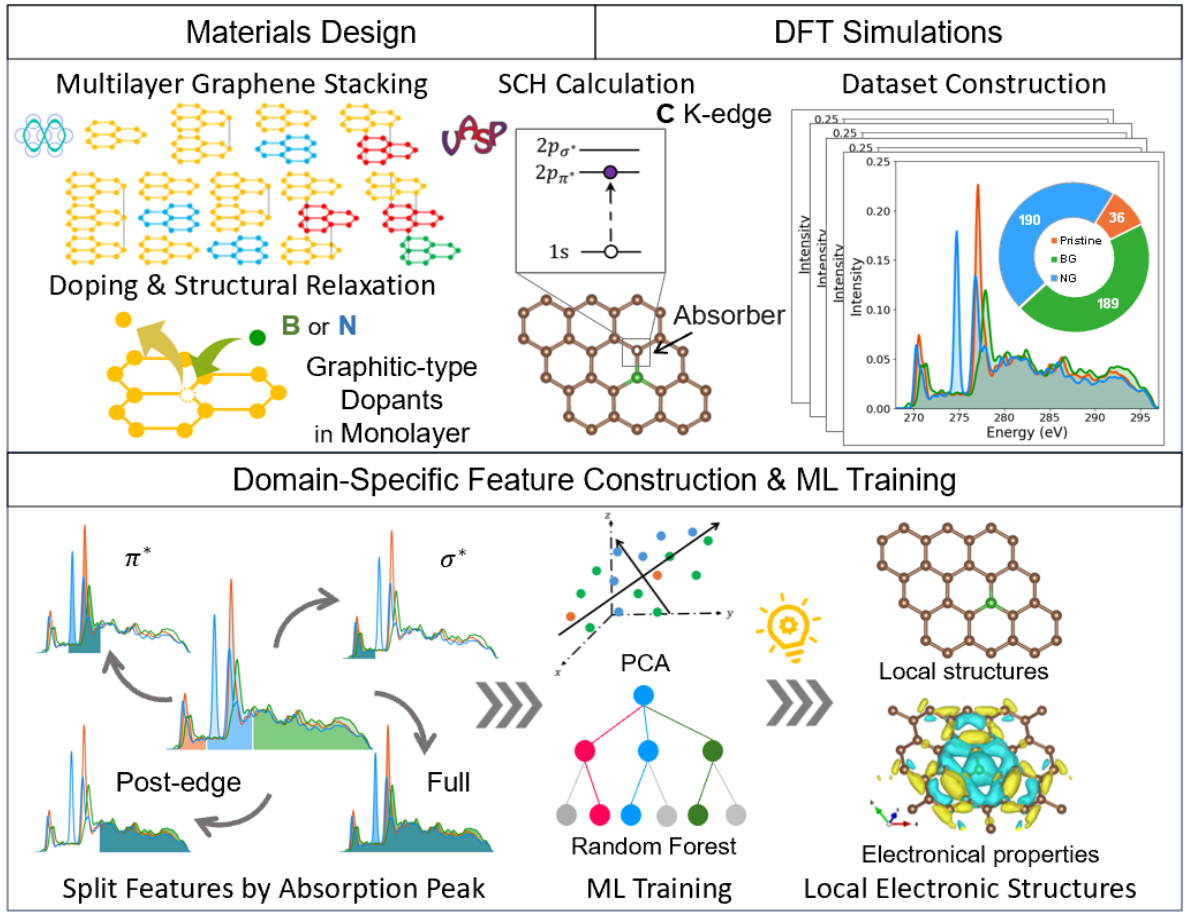

Fig.1 Comprehensive workflow for machine learning-based analysis of doped graphene X-ray absorption spectroscopy. The methodology comprises three primary stages: materials design, involving multilayer graphene stacking configurations and strategic boron/nitrogen doping followed by structural relaxation; DFT simulations, including self-consistent Hamiltonian (SCH) calculations to generate C K-edge absorption spectra and construction of a comprehensive spectral dataset; and domain-specific feature construction and machine learning training, where spectra are systematically partitioned into absorption peak regions ($\pi^*$, $\sigma^*$, post-edge, and full spectrum) to create physically meaningful feature sets. These features are subsequently analyzed using dimensionality reduction via Principal Component Analysis (PCA), followed by Random Forest (RF) classification and regression to correlate spectral signatures with local atomic structures and electronic properties, ultimately enabling electrostatic potential mapping of the material systems.

## 2. Methods

### 2.1 Materials and computational methods:

All calculations were performed using the Vienna Ab initio Simulation Package (VASP).[22–25] The exchange-correlation interactions were described by the Perdew-Burke-Ernzerhof (PBE) functional within the generalized gradient approximation (GGA).[26] The projector augmented-wave (PAW) method was employed to describe the core-valence interactions.[27] A plane-wave cutoff energy of 500 eV and the zero-damping DFT-D3 method for van der Waals dispersion corrections were applied.[28] The energy and force convergence criteria were set to $10^{-5}$ eV and 0.05 eV/Å, respectively.

The initial pristine single-layer graphene model was derived from the optimized graphite structure (mp-48) obtained from the Materials Project database. The Python library VASPKIT was used to create a 6x6x1 supercell containing 72 atoms to minimize interactions between core-holes under periodic boundary conditions.[29] In addition, a vacuum layer of 15 Å was introduced along the z-axis. Structural relaxation was performed using a Γ-centered K-point mesh of 2x2x1. Furthermore, using the Atomic Simulation Environment (ASE) library, 8 multilayer pristine graphene structures with distinct stacking configurations were constructed and fully relaxed.[30] Together with the monolayer, these 9 pristine structures consist of three bilayer variants (AA, AB, and XY) and five trilayer variants (AAA, AAB, ABA, XYX, and XYZ).

Based on the optimized single-layer graphene, boron (B) and nitrogen (N) impurities were introduced by substituting specific carbon atoms. Five different doping concentrations were investigated: 1.39%, 2.78%, 4.17%, 5.56%, and 6.95%. For the 1.39% concentration, a single B or N atom substitution was modeled. For higher concentrations, ten distinct doping configurations were generated for each to account for different spatial distributions of the dopants. In total, 82 impurity-doped and 9 pristine graphene structures were constructed and fully optimized for subsequent property calculations.

Carbon K-edge XANES spectra were simulated using the eXcited Core-hole (XCH) approach without considering spin polarization.[31,32] Core-holes were introduced specifically at inequivalent carbon sites based on their local chemical environments near the dopants, resulting in a total of 415 calculated XANES spectra. A denser Γ-centered 4x4x1 K-point mesh was employed to ensure the accuracy of the spectra. In contrast, Bader charge analyses for all atoms in each structure were performed using a Γ-centered 3x3x1 K-point mesh.

## 2.2 Machine learning methods

In this study, a comprehensive machine learning framework was applied to examine the C K-edge XANES spectra of graphene doped with boron and nitrogen. A dataset of 415 spectra served as the basis for both classification and regression analyses. To facilitate these analytical procedures, we first employed Principal Component Analysis (PCA) as an exploratory technique to understand the inherent variance structure and dimensionality of the spectral data.[33] Following this dimensional analysis, the RF model was selected for the classification and regression tasks, using the scikit-learn library (v1.6.1).[34]

All 415 spectra were first smoothed using a Gaussian filter with a standard deviation of 1.0 eV, after which the integrated area under each spectrum was normalized to unity. The initial energy point of a spectrum was defined by the onset of non-zero absorption in the dataset, and the final energy point was set by the last occurrence of non-zero absorption intensity.

To address the complex, nonlinear relationships inherent in spectral data related to dopant concentrations and crystal structures, we employed RF algorithm as our primary machine learning approach.[35] This ensemble learning method was selected based on its demonstrated capacity to handle high-dimensional spectroscopic data without overfitting, while capturing subtle spectral features that correlate with material properties. The RF methodology operates by constructing multiple decision trees during the training phase and outputting either the mode of classification trees or the mean prediction of individual regression trees, thereby providing robust predictions even in the presence of experimental factors like noise and lifetime broadening common in experimental XANES measurements.[36]

Prior to the construction of the RF model, each spectral dataset underwent systematic partitioning into five distinct energy regions: the $\pi^*$ peak window (162 features), the $\sigma^*$ peak window (158 features), the post-edge domain (422 features), and the complete spectral profile (742 features). The delineation of $\pi^*$ and $\sigma^*$ ranges was established according to spectral features observed at a nitrogen doping concentration of 1.39%, while the post-edge domain comprised energies extending beyond the absorption threshold. The analytical framework was further augmented by the incorporation of two additional spectral parameters, specifically the energetic positions corresponding to maximum absorption intensities of the $\pi^*$ and $\sigma^*$ peaks.

Supervised learning was performed by training the RF model separately on each of these five energy segments, supplemented by the two peak-position features, to identify the most informative feature domain for predicting doping concentration, mean nearest-neighbor dopant–carbon bond length, and mean dopant Bader charge. Classification was conducted independently for boron-doped and nitrogen-doped samples with doping concentrations of 0%, 1.39%, 2.78%, 4.17%, 5.56%, and 6.95%. The RF model classified each spectrum into one of these concentration categories for its respective dopant type. For the regression tasks, the model was trained to estimate the mean nearest-neighbor bond length between dopant and carbon atoms, as well as the dopant's mean Bader charge. Each of the five segmented feature sets was evaluated to determine the region most predictive of these target quantities. To optimize performance, a 5-fold cross-validation scheme was used to assess the robustness of the RF model.

### 3. Results & Discussion:

## 3.1 Machine Learning Analysis of XANES Spectra
### *3.1.1 Dimension Reduction*

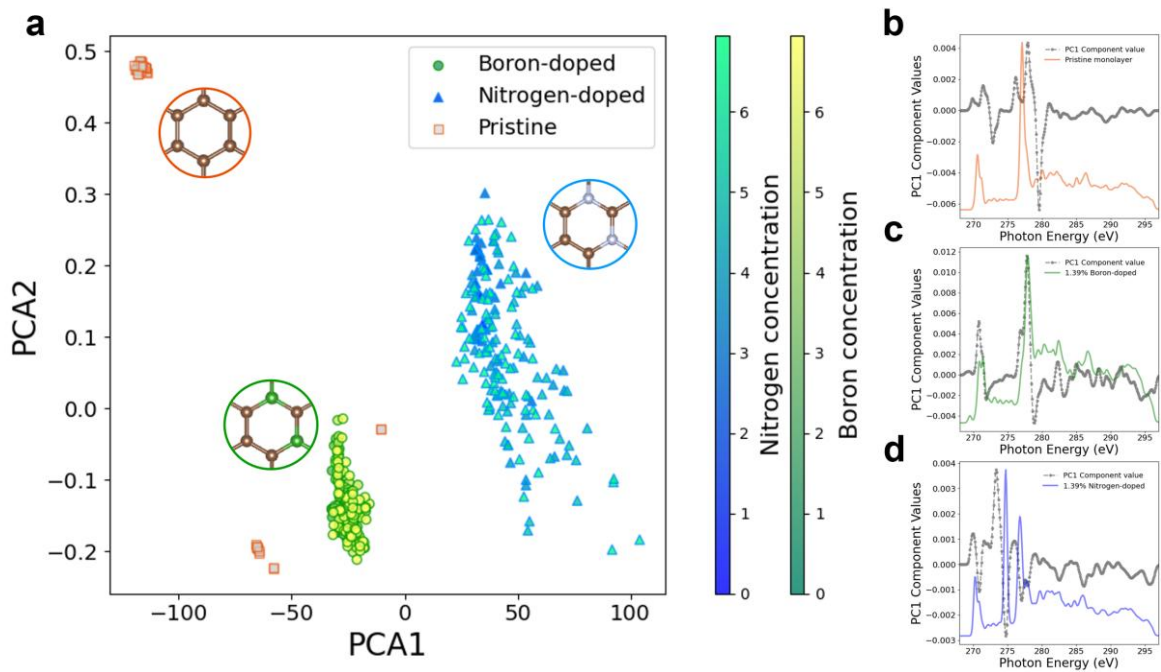

Fig. 4 Principal Component Analysis of pristine and doped graphene XANES spectra. (a) Two dimensional PCA scatter plot showing clear separation between pristine graphene (orange squares), boron-doped graphene (green circles), and nitrogen-doped graphene (blue triangles). The color intensity represents dopant concentration, with darker colors indicating higher doping levels. Structural diagrams illustrate the atomic configurations of each system. (b) PCA loading curve (gray line) with pristine monolayer graphene spectrum (orange line), highlighting spectral features that contribute most to variance. (c) PCA loading curve (gray line) compared with 1.39% doped BG spectrum (green line). (d) PCA loading curve (gray line) compared with 1.39% doped NG spectrum (blue line).

Fig. 4 presents the PCA results and corresponding loading curve, illustrating the contribution of each energy point to the first principal component. In Fig. 4(a), pristine graphene, boron-doped graphene (BG), and nitrogen-doped graphene (NG) form distinct clusters. Within the pristine graphene cluster, subgroups corresponding to mono-, bi-, and tri-layer structures are observed, reflecting layer-dependent absorption edge shifts. The BG cluster is compact, indicating high spectral similarity across boron-doped structures, whereas NG exhibits greater dispersion due to the increased electronic complexity introduced by nitrogen. Specifically, nitrogen doping promotes local sp³ hybridization near the dopant site, producing multiple σ* peaks and higher spectral variability. However, PCA does not resolve dopant concentration.

For all systems, PC1 accounts for over 99.5% of variance, suggesting that spectral differences arise predominantly from systematic energy shifts rather than localized feature modification. Consequently, PCA effectively distinguishes dopant types and graphene layer numbers but is insufficient to capture subtle structural or concentration-dependent effects. The corresponding loading curves highlight the spectral regions most sensitive to these variations, particularly the π* and σ* absorption peaks. For example, increasing the number of graphene layers shifts both the π* and σ* peaks to higher the energies, consistent with interlayer electronic coupling effects.

Analysis of the loading curves further indicates that boron doping induces a blue shift in absorption features, while nitrogen doping results in a red shift, consistent with p- and n-type Fermi level modifications, respectively. Fig. 4(c) and (d) show that varying boron concentration has minimal effect on the edge position, whereas increasing nitrogen content leads to pronounced red shifts and edge softening. These trends are attributable to hybridization changes, that is, boron largely preserves sp² bonding, whereas nitrogen introduces lone pairs and promotes partial sp³ character, lowering the energy of unoccupied states. This interpretation is further corroborated by σ* peak splitting in NG, consistent with PDOS analysis and previous experimental results.[37]

In summary, PCA successfully separates spectra by dopant type and layer number but cannot resolve finer structural or concentration details due to its linear nature. Loading curves provide a physically meaningful insight into dopant-induced spectral changes, reflecting Fermi level shifts and hybridization effects. Nonlinear dimensionality reduction methods (UMAP, t-SNE) were also tested to explore whether manifold-based approaches could uncover subtler structural distinctions. While UMAP has previously demonstrated strong capability in resolving phase variations and defect states in BN systems, in the present graphene systems both UMAP and t-SNE show similar limitations, distinguishing primarily dopant types and layer numbers.[38] This outcome likely reflects the relatively uniform structural topology of doped graphene compared to the more structurally diverse BN phases. The corresponding UMAP and t-SNE plots are provided in the Supporting Information (Fig. S1)

### 3.1.2 Classification of Doping Types

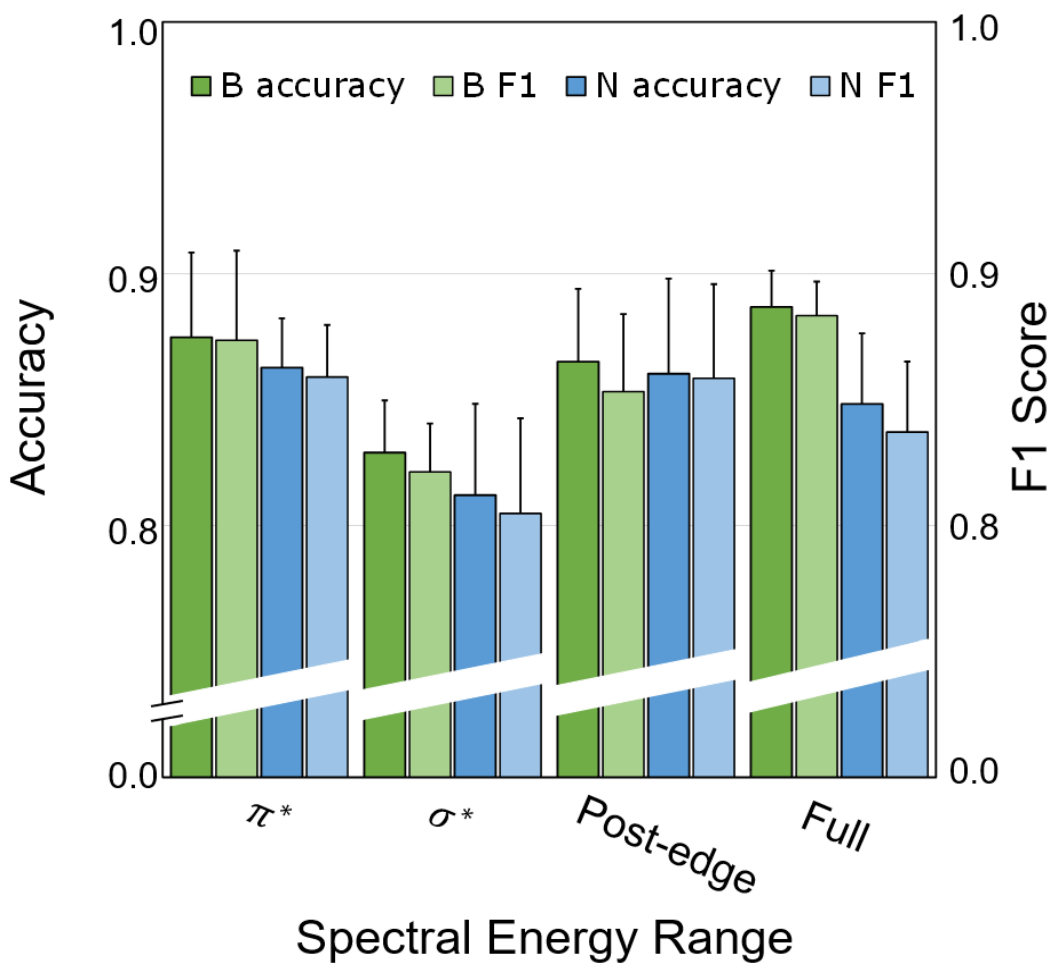

Fig. 5 Performance of RF model for dopant concentration prediction using different spectral regions. The bar chart displays the accuracy and F1 scores from 5-fold cross-validation for RF models trained to classify boron (green) and nitrogen (blue) concentrations. Performance metrics are shown for models trained using different spectral regions as input features: π* resonance, σ* resonance, post-edge region, and the full spectrum. Error bars represent ±1 standard deviation across the cross-validation folds.

Figure 5 shows that the RF model effectively captures dopant-specific signatures in the XANES spectra. The choice of spectral region has a decisive influence on classification performance. Among the four regions examined π*, σ*, post-edge, and full spectrum, the π* region provides the most balanced and robust results, achieving the highest overall accuracy and F1 scores for both boron- and nitrogen-doped systems. Other regions exhibit lower performance and greater variability, reflecting a weaker correlation with dopant-induced spectral changes. Five-fold cross-validation further confirms the robustness of the π* region, showing minimal standard deviation and negligible overfitting across all test folds. It should be emphasized that the RF model identifies statistical patterns within the spectra but does not itself explain the underlying physics. The physical origin of the classification performance must therefore be interpreted in terms of electronic structure modifications induced by doping.

Pristine graphene has an electronic structure made up of σ and π bonds that originate from sp² hybridization. The π* orbitals are the frontier orbitals and define the density of states near the Fermi level, while the σ* orbitals lie at higher energies. When a carbon atom is replaced by boron or nitrogen, several types of charge rearrangement occur. First, local Friedel oscillations redistribute charge where boron atoms donate electrons to nearby carbon atoms, whereas nitrogen atoms withdraw electron density. Second, doping modifies bond lengths and twists bond angles, partially breaking the planar symmetry of the sp² network and introducing lattice strain. The $p_z$ orbitals of the dopant and its neighboring carbon atoms become perturbed, new hybridized states form, the π network loses partial delocalization, and localized defect states emerge. This behavior aligns with the findings reported by Wang *et al.*, who showed that introducing heteroatoms generates defect sites that significantly alter the electronic structure and surface properties of graphene.[39] As the dopant concentration increases, the π * resonance correspondingly broadens. Third, the σ system also affected. Difference charge density maps (Fig.S3) reveal pronounced polarization within σ bonds around ~~of~~ the absorbing atoms, indicating substantial in-plane electronic rearrangements. The π* peak resonance reflects the occupancy, spatial distribution, and band structure of the π* states.[40] Since doping-induced charge reconstruction primarily modifies the frontier electronic states, changes first appear in the π* region. With increasing dopant concentration, the local density of unoccupied states, particularly in the conduction band evolves more significantly, which is clearly reflected in the spectral features. Usachov *et al*. experimentally observed concentration-dependent local electronic states in doped graphene systems. Their XANES measurements of boron-doped graphene demonstrated that doping initially reconstructs the π* characteristics, and with increasing dopant content, the conduction-band local state density undergoes more pronounced modification.[11] These experimental observations are consistent with the spectral trends identified in the present analysis.

Overall, these results indicate that the π* XANES region constitutes a physically meaningful and chemically sensitive feature space for ML-based classification. The RF model successfully exploits correlation between π*-related spectral features and dopant-induced electronic structure modifications. However, the interpretability arises from the underlying electronic structure changes; namely hybridization perturbation, charge redistribution, and defect-state formation, rather than from the ML algorithm itself. Detailed quantitative metrics, cumulative importance indices, and comparison with previous experimental and theoretical reports are provided in the Supplementary Information (Table S1).

### 3.1.3 Regression of structural and electronic properties

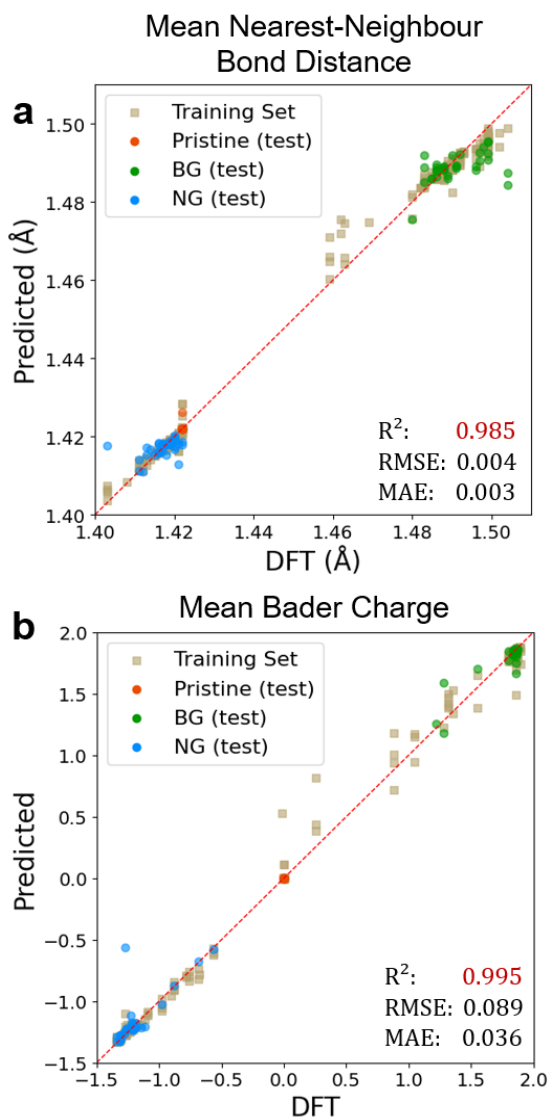

Fig. 6 Parity plots by RF models comparing predicted and DFT-calculated properties of doped graphene using the π* region. (a) Mean nearest-neighbor distances (Å) between dopants and adjacent carbon atoms. (b) Mean Bader charge of dopants. Predicted values (y-axis) are plotted against DFT-calculated values (x-axis). Beige squares indicate training set points, while orange, green, and blue circles correspond to pristine graphene, boron-doped graphene (BG), and nitrogen-doped graphene (NG), respectively. Both plots demonstrate the high predictive accuracy of the π* model for structural and electronic descriptors.

We achieved strong regression performance for both the C-dopant bond distances and Bader charges, as illustrated in Fig. 6. The parity plots display strong linear correlations between predicted and DFT-calculated values for both training and test sets, indicating that the RF model captures statistically meaningful relationships between XANES spectral features and atomic-scale properties.

For bond-distance prediction, the π* energy region provided the highest overall accuracy. Quantitatively,

the $\pi^*$ model reduced the RMSE by 4.35 %, 6.38 %, and 8.33 % compared with the $\sigma^*$, post-edge, and full-spectrum models, respectively (details in SI). The MAE also decreased by 7-10% relative to other regions. Cross-validation further confirmed that the $\pi^*$ window yielded the lowest standard deviation in $R^2$, indicating stable and reproducible learning behavior. In contrast, the $\sigma^*$ and post-edge regions introduced greater spectral redundancy and weaker sensitivity to local bonding perturbations, leading to slightly reduced predictive performance.

The improved regression accuracy of the $\pi^*$ region can be rationalized by its intrinsic sensitivity to local charge redistribution and orbital hybridization near the dopant site. Substitutional boron or nitrogen induces charge transfer between the dopant and neighboring carbon atoms, accompanied by local lattice distortion, $p_z$ orbitals tilting, and partial $sp^3$ rehybridization.[39] These structural and electronic perturbations directly influence the $\pi^*$ transition energy and intensity, thereby establishing a strong statistical correlation between $\pi^*$-region spectral features and C-dopant bond distances. Experimental studies have similarly shown that perturbations in the $\pi$-band constitute the earliest and most pronounced spectral signatures of dopant incorporation in graphene. [11,39]

For charge prediction, the $\pi^*$ region again yielded the most reliable performance. The RF model trained on $\pi^*$ spectra achieved $R^2$ = 0.9952, RMSE = 0.0968 $e^-$, and MAE = 0.0512 $e^-$, outperforming the $\sigma^*$ and post-edge models. Although the full-spectrum model achieved a comparable cross-validation $R^2$, its test-set errors suggest reduced generalization relative to the $\pi^*$-restricted model (see Fig. S4).

This behavior can be understood from the fundamental electronic structure of graphene. In the $sp^2$ network, $\pi$ electrons and $\sigma$ electrons are largely decoupled in energy and function.[41] The $\sigma$ electrons form deep, localized bonding states located well below the Fermi level, and primarily determine lattice stability. Extensive theoretical work shows that these $\sigma$ states exhibit strong chemical rigidity and limited participation in low-energy charge redistribution, even under heteroatom substitution.[42]

In contrast, the $\pi$-electron system, composed of delocalized $p_z$ orbitals forming $\pi$ and $\pi^*$ bands intersect the Fermi level, governs the low-energy electronic response of graphene.[41,43] Both experimental and theoretical studies confirm that substitutional dopants primarily perturb the $\pi$ manifold, while the $\sigma$ backbone remains comparatively inert.[44]

Boron, possessing one fewer valence electron than carbon, acts as a $\pi$-electron acceptor. It withdraws electron density from neighboring π orbitals, generating localized hole states and shifting the unoccupied π* states to higher energy.[43] Nitrogen produces the opposite effect, i.e., its additional valence electron preferentially populates π states, modifying the distribution of unoccupied π* orbitals while largely preserving σ bonding.[13,40,45] Spatial charge-density difference maps and Bader analyses confirm that these redistributions are concentrated within the π network.

Because the Bader charge quantifies ground-state electron density redistribution, and this redistribution occurs predominantly in π orbitals, spectral features probing π states naturally become the most informative descriptors. The C K-edge π* resonance, arising from $1s \rightarrow \pi^*$ transitions, directly probes the unoccupied π-electron density around carbon atom. Dopant-induced changes in π* onset energy, intensity, and spectral shape therefore correlate strongly with local electron enrichment or depletion.[45] Conversely, the σ* region probes $1s \rightarrow sp^2$ σ antibonding states that remain comparatively stable under substitution, limiting their sensitivity to charge transfer.

~~VASP~~

Within this context, the superior performance of the π* window does not arise from the ML algorithm itself, but from the fact that π* features encode the physically relevant degrees of freedom governing dopant-induced electronic and structural perturbations. The RF model identifies statistical correlations embedded in these spectroscopic signatures, while the underlying interpretability is grounded in well-established electronic-structure principles of doped graphene.

**Conclusion:**

In this work, we developed an interpretable ML framework to analyze XANES spectra of doped graphene by separating contributions from physically distinct spectral regions. Our findings reveal that the π* region yields the most reliable predictive performance for local structural and electronic descriptors, including Bader charge and C-dopant bond length, compared with σ*, post-edge, and full-spectrum approaches. The effectiveness of the π* resonance stems from its intrinsic sensitivity to charge redistribution and orbital hybridization within the π-electron network perturbed by boron or nitrogen dopants.

By integrating DFT-simulated datasets with RF modeling, we demonstrate that region-resolved spectral decomposition enhances predictive stability while preserving physical interpretability. Rather than relying on the full spectrum as a black-box input, isolating physically meaningful spectral windows allows the model to focus on the degrees of freedom most relevant to dopant-induced electronic perturbations. Although demonstrated here for doped graphene, the proposed framework provides a transferable strategy for analyzing spectroscopic datasets in systems where distinct spectral regions correspond to separable physical processes. This approach may therefore be extended to other two-dimensional materials and heteroatom-modified carbon systems, provided that appropriate physical insight guides feature selection.

Overall, this study highlights the importance of combining domain knowledge with machine learning to ensure that predictive performance remains grounded in established electronic-structure principles. Region-resolved, physically motivated learning offers a pathway toward more transparent data-driven spectroscopy and improved integration between computational modeling and experimental characterization of advanced carbon-based materials.

## Data availability
The data used in this study are available from the corresponding authors upon reasonable request.

## Code availability
The codes used in this study are available from the corresponding authors upon reasonable request.

## Declaration of competing interest
The authors declare no competing interests.

## Acknowledgements
This work was funded by Japan Science and Technology Agency (JST) CREST (Grant No. JPMJCR21O4). AV and MK thank Research Center for Computational Science, Okazaki, Japan for supercomputing facilities received for some calculations (Project: 24-IMS-C132). This work was fully conducted using the facilities provided by the Tokyo University of science and YW is now affiliated with The University of Tokyo. YW acknowledges support from the Program for Leading Graduate Schools (MERIT-WINGS) of the University of Tokyo. The authors thank Dr. Alexandre Lira Foggiatto from the Tokyo University of Science and Zengqing Wu from the University of Osaka for useful discussions on machine learning.

**TOC:**

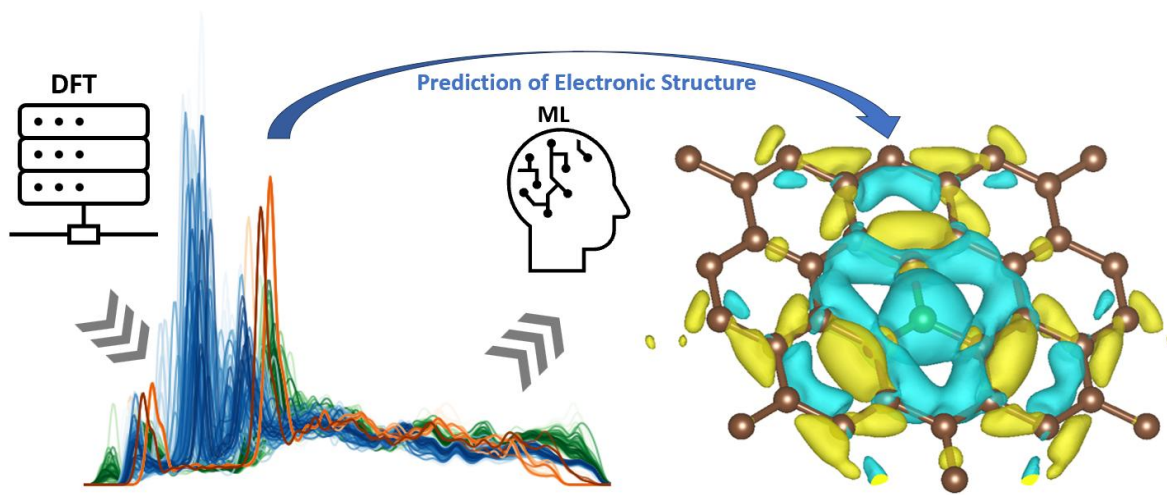

# Decoding Dopant-Induced Electronic Modulation in Graphene via Region-Resolved Machine Learning of XANES

Yinan Wang[1,2,3], Arpita Varadwaj[1*], Teruyasu Mizoguchi[2], Masato Kotsugi[1]

[1] Department of Material Science and Technology, Tokyo University of Science, 6-3-1, Niijuku, Katsushika, Tokyo, 125-8585, Japan

[2] Institute of Industrial Science, The University of Tokyo, 4-6-1, Komaba, Meguro, Tokyo 153-8505, Japan

[3] Department of Materials Engineering, the University of Tokyo, 7-3-1, Hongo, Bunkyo, Tokyo, 113-8656, Japan

*Corresponding author's e-mail: varadwaj.arpita@gmail.com

**Electronic Supplementary Information (ESI)**

## S1. Machine Learning Evaluation Metrics

Classification performance was quantified by accuracy and F1 scores, while regression outcomes were measured using the coefficient of determination ($R^2$), mean absolute error (MAE), and root mean squared error (RMSE). Their definitions are shown in equations (3) to (6), where $y_i$ is Actual value, $\hat{y_i}$ is predicted value, $\bar{y}$ is the mean of $y_i$, $n$ is the number of data points.

$$\text{Precision} = \frac{\text{True Positives (TP)}}{\text{True Positives(TP)} + \text{Flase Positives (FP)}} \quad (1)$$

$$\text{Recall } = \frac{\text{True Positives (TP)}}{\text{True Positives(TP) } + \text{ Flase Negatives (FN)}} \tag{2}$$

$$\text{F1 score} = 2 \times \frac{\text{Precision } \times \text{ Recall}}{\text{Precision } + \text{ Recall}} \tag{3}$$

$$R^2 = 1 - \frac{\sum_{i=1}^{n}(y_i - \widehat{y_i})^2}{\sum_{i=1}^{n}(y_i - \bar{y})^2} \tag{4}$$

$$\text{MAE } = \frac{1}{n}\sum_{i=1}^{n}|y_i - \widehat{y_i}| \tag{5}$$

$$\text{RMSE } = \sqrt{\frac{1}{n}\sum_{i=1}^{n}(y_i - \widehat{y_i})^2} \tag{6}$$

## S2. UMAP and t-SNE

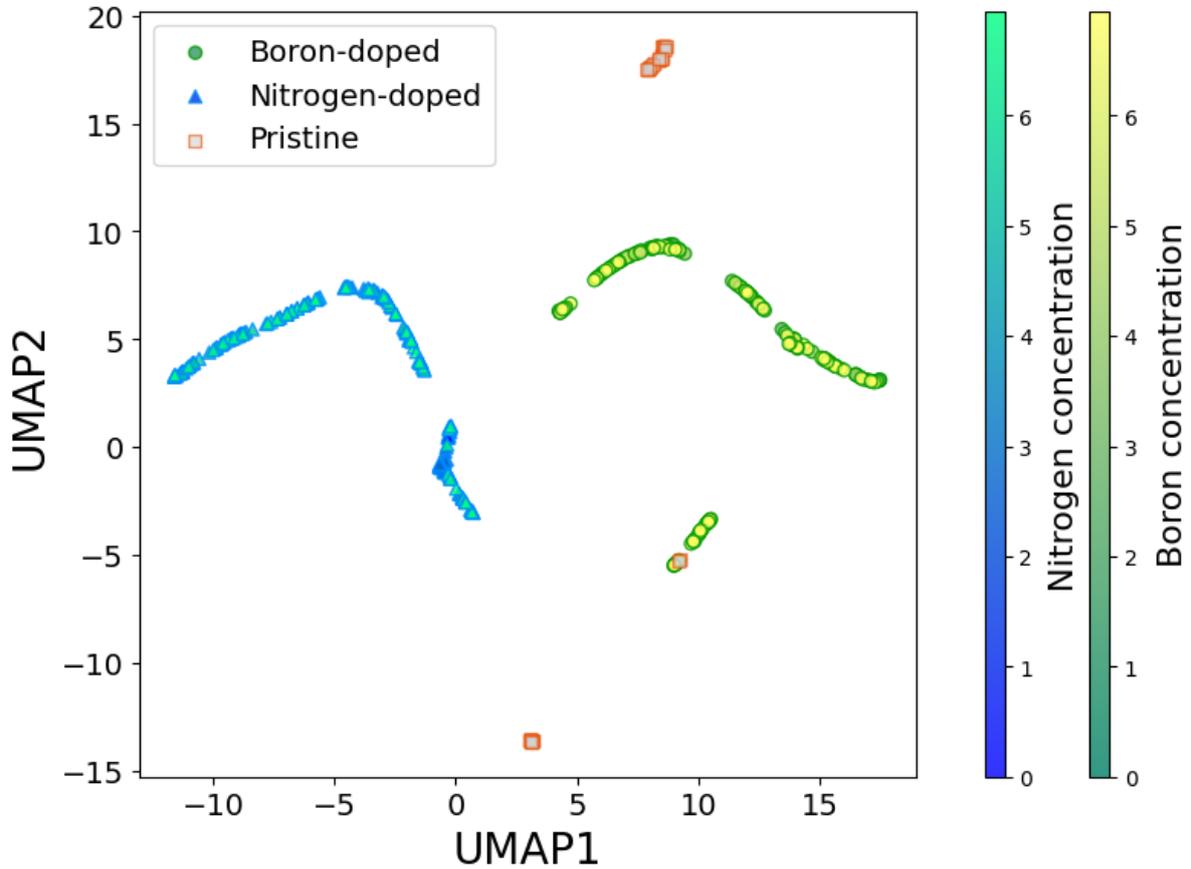

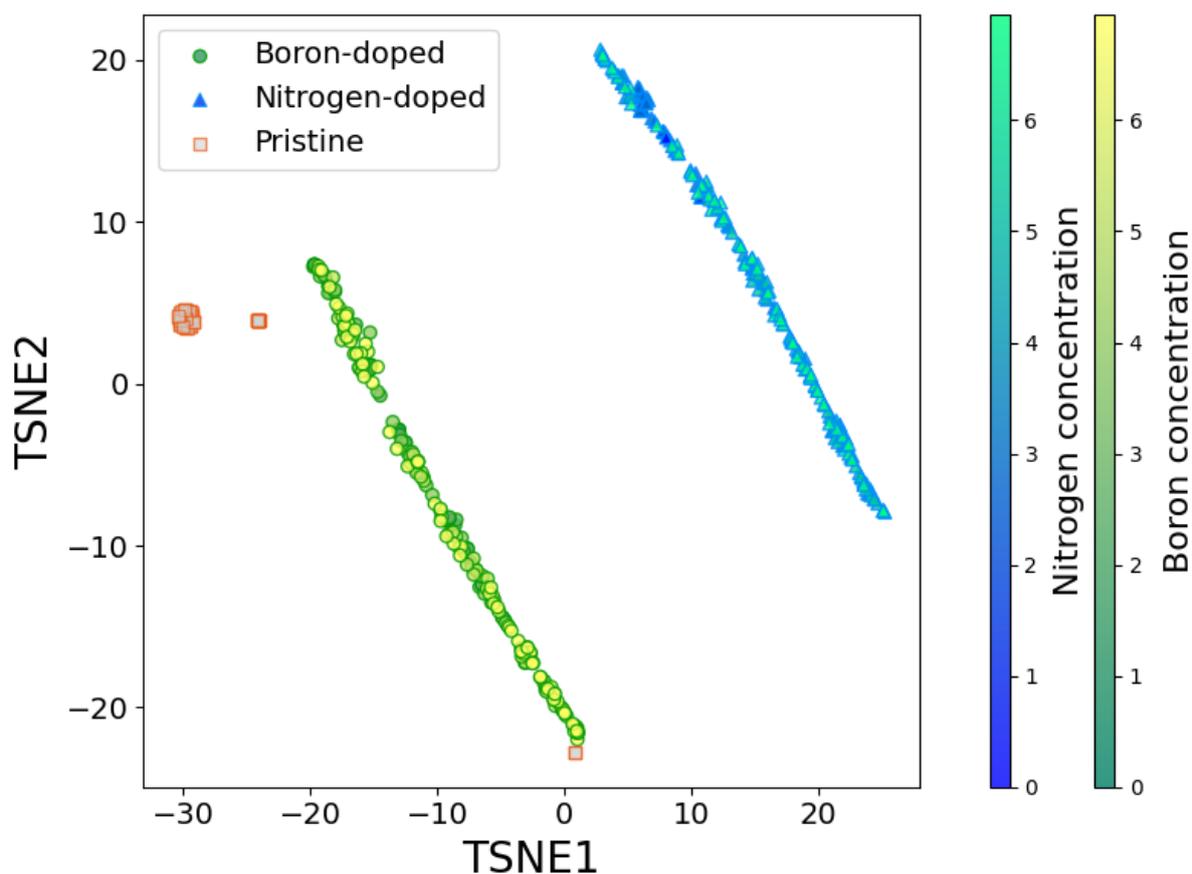

## S3. Classification Metrics for Dopant Types

### S3.1 Random Forest Performance for Different Spectral Regions

Table S1 summarizes the 5-fold cross-validation results for the random forest model trained on four spectral regions: π*, σ*, post-edge, and full spectrum. The metrics include class-wise accuracy, F1 scores, macro-averaged metrics, and standard deviations.

**Table S1.** Cross-validation metrics for RF classification of boron- and nitrogen-doped graphene.

| Spectral Region | Accuracy (Boron) | Accuracy (Nitrogen) | F1 Score (Boron) | F1 Score (Nitrogen) | Std. Dev. (Accuracy) | Std. Dev. (F1) |
|---|---|---|---|---|---|---|
| π* Region | 0.985 | 0.973 | 0.984 | 0.969 | 0.007 | 0.008 |
| σ* Region | 0.956 | 0.938 | 0.954 | 0.935 | 0.012 | 0.015 |
| Post-Edge | 0.962 | 0.940 | 0.960 | 0.937 | 0.011 | 0.013 |
| Full Spectrum | 0.989 | 0.955 | 0.987 | 0.949 | 0.009 | 0.012 |

**S4. Regression Performance and Model Evaluation**

**Table S2.** Performance metrics of random forest models trained on different XANES spectral regions for predicting Bader charge (e⁻).

| Spectral Region | $R^2$ | RMSE (e⁻) | MAE (e⁻) |
|---|---|---|---|
| π* Region | 0.9952 | 0.0968 | 0.0512 |
| σ* Region | – | 0.236 | 0.123 |
| Post-Edge Region | – | 0.163 | 0.093 |
| Full Spectrum | – | 0.098 | 0.052 |

The π* region yields the best regression performance with the highest $R^2$ (0.9952) and the lowest RMSE (0.0968 e⁻), demonstrating its superior sensitivity to local charge redistribution and electronic perturbations around dopants.

Models trained on σ* and Post-Edge regions show higher prediction errors, indicating weaker spectral correlation with local Bader charge variations.

The Full spectrum slightly improves prediction for boron-doped systems due to enhanced π* contributions, but introduces spectral noise that limits generalization for nitrogen-doped systems.

**Table S3.** Performance metrics of random forest models trained on different XANES spectral regions for predicting mean nearest-neighbor (NN) distance (Å).

| Spectral Region | $R^2$ | RMSE (Å) | MAE (Å) |
|---|---|---|---|
| π* Region | 0.992 | 0.011 | 0.006 |
| σ* Region | 0.973 | 0.024 | 0.013 |
| Post-Edge Region | 0.955 | 0.031 | 0.016 |
| Full Spectrum | 0.987 | 0.018 | 0.010 |

**S5. Spectral features analysis of Pristine graphene and its doped analogues**

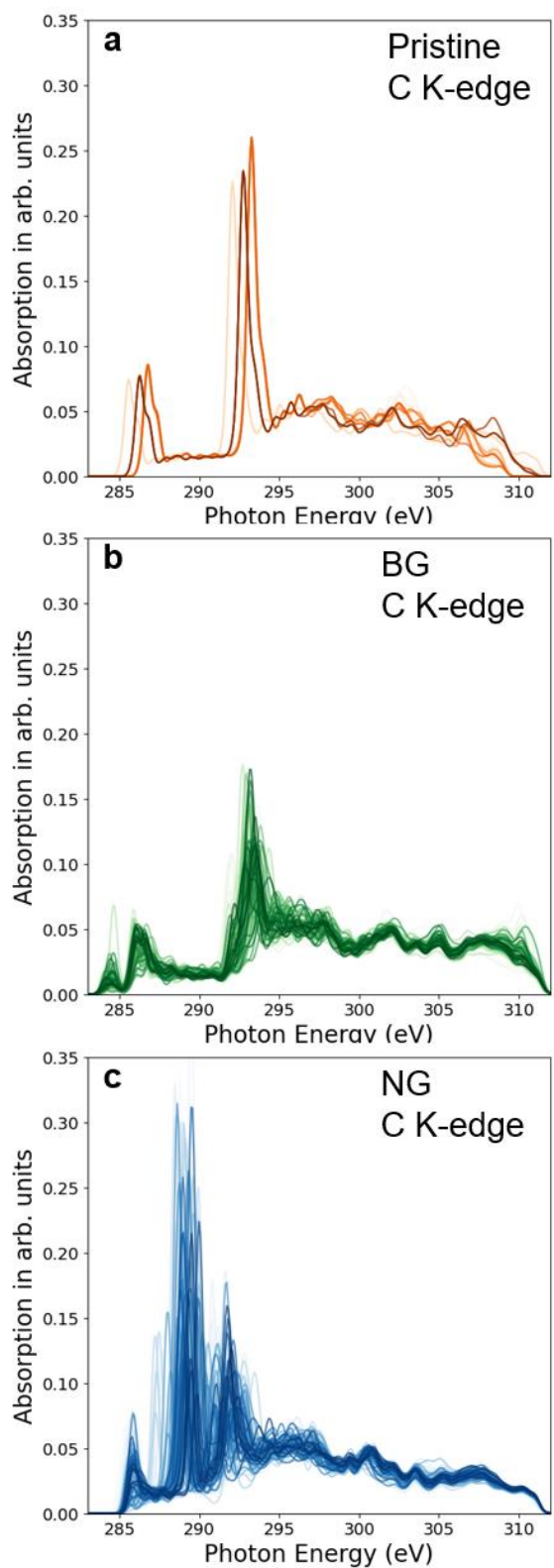

Fig. S2 Carbon K-edge X-ray Absorption Near Edge Structure (XANES) spectra calculated using PBE-GGA. (a) Pristine graphene from monolayer to trilayer with different stacking. (b) Boron-doped graphene (BG) with doping concentrations ranging from 1.39% to 6.95%, presented as a heat map overlay of multiple spectra. (c) Nitrogen-doped graphene (NG) with doping concentrations ranging

from 1.39% to 6.95%, also presented as a heat map overlay of multiple spectra. The spectra reveal distinct electronic structure modifications induced by heteroatom doping compared to pristine graphene, particularly in the 285-295 eV energy range where $\pi^*$ and $\sigma^*$ transitions occur.

For pristine graphene, we observe two well-separated resonances, corresponding to transitions to anti-bonding orbitals. These distinct spectral features reflect electron excitations from the C 1s core level to unoccupied $\pi^*$ and $\sigma^*$ anti-bonding states. Our theoretical spectra and excited-state PDOS calculations for pristine graphene reveal the absence of pre-edge features, while the $\pi^*$ resonance manifests as a characteristic doublet structure. These findings are consistent with the theoretical XANES results reported by Hua et al., who demonstrated that this doublet does not originate from two distinct $\pi^*$ states but rather from the superposition of several intrinsic orbital transitions. Both our work and theirs attribute this spectral feature to finite supercell size effects, confirming that computational artifacts from periodic boundary conditions must be carefully considered when interpreting theoretical carbon K-edge spectra of graphene systems.[1,2]

In multilayer graphene, spectra taken with the core hole placed on different carbon atoms within the same layer are identical. Moreover, regardless of the interlayer twist angle or the stacking sequence, the $\pi^*$ and $\sigma^*$ features of each carbon remain almost unchanged, showing that the electronic states are essentially uniform in pristine graphene. Clear differences arise only in the post-edge region. Relative to the monolayer, the main absorption peak of a bilayer is blue-shifted by about 1.0 eV, and that of a trilayer by about 1.7 eV, again consistent with Hua et al. [2]

Doping graphene with boron or nitrogen strongly perturbs its $\pi$-electron system, as reflected in the $\pi^*$ region of the C K-edge XANES spectra (Fig. 2b,c). In boron-doped graphene (BG), the $\pi^*$ peak exhibits a clear blue shift and splitting, indicating electron deficiency at the boron sites and local structural distortion. A distinct pre-edge feature appears along the z-direction, associated with partially unoccupied $\pi$ orbitals resulting from charge depletion near boron. Projected density of states (PDOS) analysis confirms that the $\pi^*$ resonance mainly originates from the $p_z$ orbitals of carbon atoms bonded to boron, while the $\sigma^*$ region retains primarily $sp^2$ hybridization with minor s, $p_x$, and $p_y$ contributions. These features signify partial $sp^3$ character around the boron dopant, driven by out-of-plane distortions and charge redistribution.

In contrast, nitrogen-doped graphene (NG) exhibits a weakened $\pi^*$ peak and a distinct splitting in the $\sigma^*$ region, reflecting partial $sp^3$ hybridization. PDOS analysis shows that the conduction states in NG are dominated by $p_x$ and $p_y$ orbitals of nitrogen and its neighboring carbon atoms, with small s-orbital contributions and negligible $p_z$ involvement [45,50]. This indicates that nitrogen largely preserves planar $sp^2$ bonding while introducing mild electronic rearrangements through lone-pair interactions and localized charge transfer. Consequently, boron acts as an electron acceptor, creating hole-like regions, whereas nitrogen serves as an electron donor, producing electron-rich sites. Notably, the $\pi^*$ region captures these dopant-induced electronic perturbations most directly, explaining its high predictive power in the machine-learning framework.

The evolution of the C K-edge spectra with dopant concentration (1.39 %–6.95 %) further supports these interpretations. In BG, the $\pi^*$ splitting and pre-edge intensity remain qualitatively similar across all concentrations, whereas $\sigma^*$ broadening slightly increases with higher boron content. In NG, the weakened $\pi^*$ intensity and $\sigma^*$ splitting persist with increasing nitrogen concentration, indicating that the dopant-induced perturbations are robust over the studied range.

The pre-edge feature observed in BG corresponds to the 1s → 2pz transitions of carbon atoms adjacent to boron.[1,3] Minor variations in pre-edge intensity arise from finite supercell effects and subtle local distortions. For NG, the absence of a pre-edge peak is consistent with the nearly full occupancy of the $\pi$ orbitals at the nitrogen site. The small $\pi^*$ splitting in BG and subtle $\sigma^*$ changes in both doped systems can be partially attributed to finite supercell size in the DFT simulations.[2] These computational artifacts, however, do not affect the main spectral trends and reinforce the interpretation that dopant-induced modifications of the $\pi$ electronic network dominate the observed spectral evolution.

**S6. Charge Difference maps.**

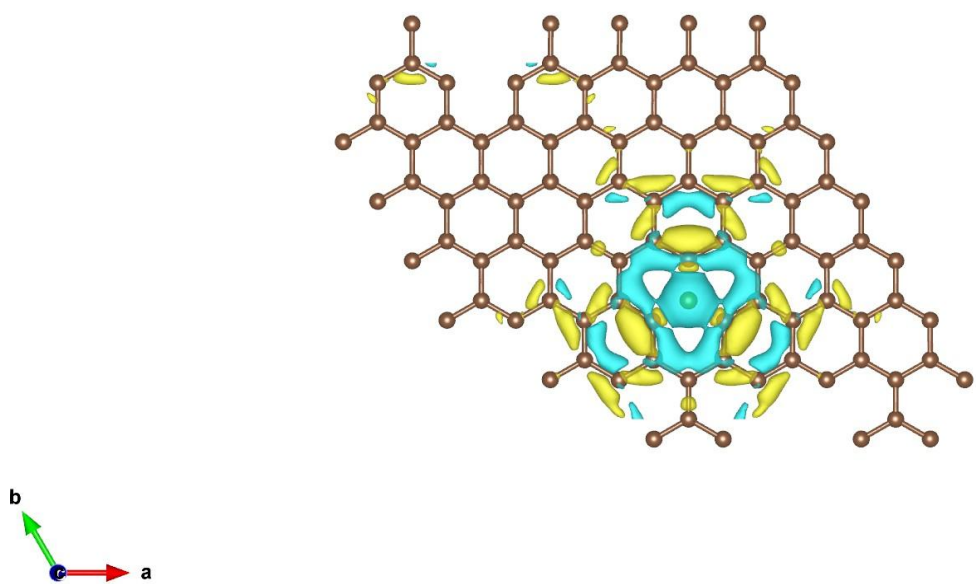

Fig.S3 Charge density difference of the 1.39% B-doped graphene supercell. The cyan and yellow isosurfaces represent positive (electron accumulation) and negative (electron depletion) charge densities, respectively.

Higher dopant concentrations, especially in NG, add extra holes or electrons. The new states give extra absorption features, so the σ* resonance in the C K-edge XANES develops many small ripples and becomes much wider. Because of this complexity, the σ* region provides too little clear information, and the weak performance of the machine-learning model on that region confirms the point, the excess variation in the σ * region confuses the model and makes it hard to capture enough signal.

**S7. Relationship between σ*–π* energy separation and the C–C bond length**

Previous work by Rojas et al. reported a linear correlation between the σ*–π* energy separation and the C–C bond length in rippled graphene.[4] However, this relationship breaks down in doped systems. This discrepancy arises because all atoms remain nearly coplanar, minimizing σ* variations, while dopant-induced local distortions dominate the π* region.

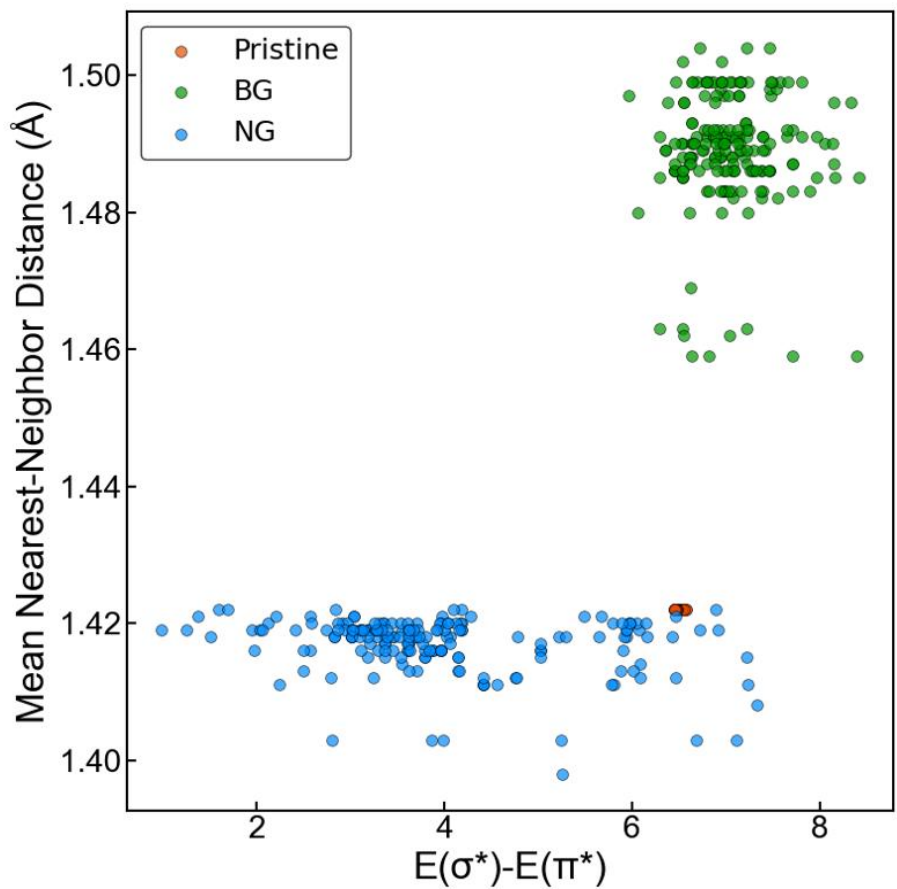

Fig.S4 Relationship between the σ*-π* energy separation and the mean nearest-neighbor distance for pristine, BG, and NG.